\documentclass{article}

\usepackage[utf8]{inputenc}
\usepackage{authblk}
\usepackage{natbib}
\usepackage{comment}
\usepackage{pdfpages}
\usepackage{amssymb}
\usepackage{bbold}
\usepackage{amsmath}
\usepackage{graphicx}
\usepackage[english]{babel}
\usepackage{amsmath}
\usepackage[T1]{fontenc}
\usepackage[titletoc, title]{appendix}
\usepackage[ruled,vlined]{algorithm2e}
\usepackage{enumitem}
\usepackage{multirow}
\usepackage{lscape}
\usepackage{caption}
\usepackage{appendix}
\usepackage{subcaption}
\usepackage[noend]{algpseudocode}
\usepackage{soul}
\usepackage{xr-hyper}
\usepackage[urlcolor=cyan,   
                linkcolor=black,
                citecolor=blue,
                colorlinks=true]{hyperref}
\usepackage[left=2.5cm,right=2.5cm,top=2.5cm,bottom=2.5cm]{geometry}
\usepackage{xcolor}
\definecolor{mycol}{RGB}{25,25,25}
\definecolor{point}{RGB}{0,0,139}
\definecolor{myyellow}{RGB}{249, 166, 2}
\definecolor{mypurple}{RGB}{81, 30, 95}
\definecolor{mygreen}{RGB}{34, 144, 140}

\usepackage{graphicx}
\makeatletter
\newcommand*{\addFileDependency}[1]{
  \typeout{(#1)}
  \@addtofilelist{#1}
  \IfFileExists{#1}{}{\typeout{No file #1.}}
}
\makeatother

\definecolor{black}{RGB}{0,0,0}

\newcommand{\matrice}[1]{\boldsymbol{#1}}

\title{Post-clustering difference testing: valid inference and practical considerations}

\author[,1,2,3]{Benjamin Hivert\thanks{Corresponding author: \texttt{benjamin.hivert@u-bordeaux.fr}}}
\author[4]{Denis Agniel}
\author[1,2,3,5]{Rodolphe Thiébaut}
\author[1,2,3]{Boris P Hejblum}

\affil[1]{Univ. Bordeaux, Inserm Bordeaux Population Health Research Center, SISTM team, UMR 1219,
	Bordeaux F33076, France}
\affil[2]{INRIA Bordeaux Sud Ouest, SISTM team Talence F-33400, France}
\affil[3]{Vaccine Research Institute, VRI, Hôpital Henri Mondor, Créteil F-94000, France}
\affil[4]{Rand Corporation, Santa Monica, CA 90401, USA}
\affil[5]{CHU Pellegrin, Groupe Hospitalier Pellegrin, Bordeaux F-33076, France}

\begin{document}

\maketitle

\begin{abstract} 

Clustering is part of unsupervised analysis methods that consist in grouping samples into homogeneous and separate subgroups of observations also called clusters. To interpret the clusters, statistical hypothesis testing is often used to infer the variables that significantly separate the estimated clusters from each other. However, data-driven hypotheses are considered for the inference process, since the hypotheses are derived from the clustering results. This double use of the data leads traditional hypothesis test to fail to control the Type I error rate particularly because of uncertainty in the clustering process and the potential artificial differences it could create. We propose three novel statistical hypothesis tests which account for the clustering process. Our tests efficiently control the Type I error rate by identifying only variables that contain a true signal separating groups of observations.

\textbf{Key words: } Clustering, hypothesis testing, double-dipping, circular analysis, selective inference, multimodality test, Dip Test
\end{abstract}

\newpage

\section{Introduction}

Cluster analysis is ubiquitous in medical research (see \citet{mclachlan1992cluster} for a comprehensive overview) to perform data classification, data exploration, and hypothesis generation \citep{xu2008clustering}. Clustering works by grouping homogeneous observations into disjoint subgroups or clusters. When multivariate data are clustered, it is common to seek to identify which variables distinguish two or more of the estimated clusters, in order to interpret the clustering structure and characterise observation groups and how they differ from each other.

Despite the widespread use of clustering, \citet{hennig2015handbook} state there is no commonly accepted and formal definition of what clusters are. In fact, the definition of what a cluster should be varies depending on the context and the analysis specifics. Here we will use the definition from \citet{everitt2006handbook}, which includes only two criteria: i) homogeneity of observations within a cluster and ii) separability of observations between two different clusters. These two criteria are general enough to encompass the majority of the working definitions of clusters. 
Both can be quantified using various approaches such as distances or similarity metrics, shape of distribution \citep{steinbach2004challenges}, multimodality \citep{kalogeratos2012dip, siffer2018your}, or distributional assumptions\citep{liu2008statistical, kimes2017statistical}.

\begin{figure} \centering \includegraphics[width = .9\textwidth]{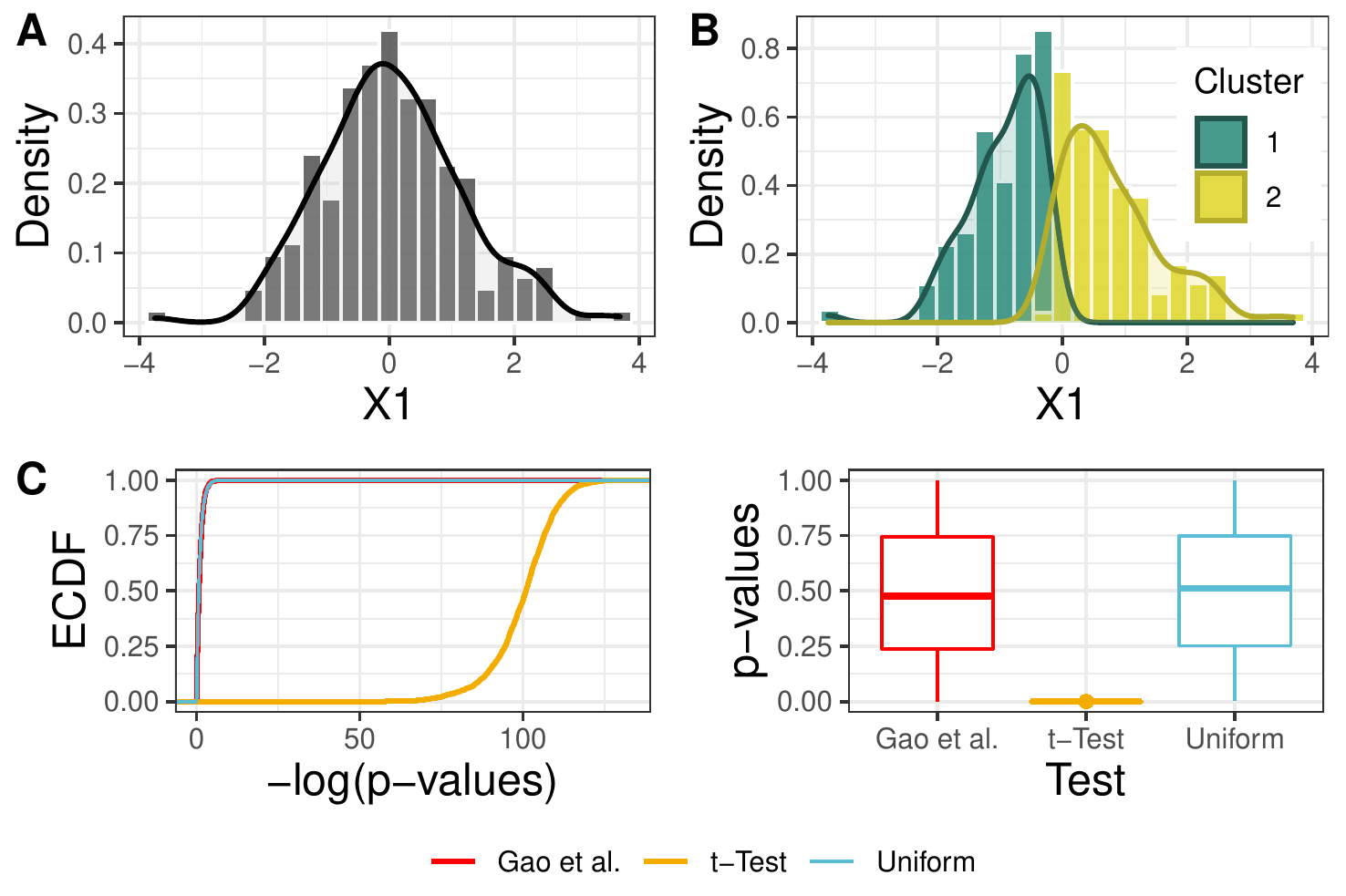} \caption{Artificial differences created by clustering. \textbf{panel A} Data generated according to $200$ realisations of a Gaussian distribution with mean $0$ and variance $1$. \textbf{panel B} Hierarchical clustering with Ward method and Euclidean distance is applied to build two clusters. \textbf{panel C} t-test p-values and p-values given by the test proposed by \citet{gao2020selective} for separating the two estimated clusters. The uniform distribution is also shown for comparison} \label{fig:1}
\end{figure}

While clustering is a multivariate methodology that takes into account all variables, only a set of variables can be expected to differentiate two particular clusters (i.e. separate their observations, according to the second criterion of our definition above). This question of which variable separate clusters of individuals is particularly relevant for high-dimensional data such as omics data \citep{ntranos2019discriminative, vandenbon2020clustering}. The current practice to identify such variables is often based on post-clustering hypothesis testing. It leads to a two-step pipeline (a first step of clustering and a second step of inference) that is actually testing data-driven hypotheses in a process sometimes referred to as ``double dipping''\citep{kriegeskorte2009circular}. This approach does not efficiently control the type I error rate when testing for differences between clusters. In fact, it is always possible to cluster the data using a clustering method, even if there is no real process separating groups of observations. In this case, the clustering method artificially enforces the differences between the observations by dividing them into different clusters. The significant differences between clusters identified during the inference process could just be an artifact of the previous clustering step. To illustrate this phenomenon, we consider data generated from a univariate Gaussian distribution with mean $0$ and variance $1$ (Figure \ref{fig:1} \textbf{panel A}). Two clusters can be built, e.g., using hierarchical clustering with Ward's method and Euclidean distance (Figure \ref{fig:1} \textbf{panel B}). These two estimated clusters are not separated clusters, since all observations come from the same Gaussian distribution. One way to infer their separation is to test for a mean shift between them, for example using the classical t-test. Since there is no real process separating these two clusters, the resulting p-values should be uniformly distributed. However, when we look at the p-values of the t-test for $2000$ simulations of the data, the resulting p-values are too small, leading to false positives (Figure \ref{fig:1} \textbf{panel B}). This simple example illustrates how it is possible to infer a separation of two clusters, even if this separation is not explained by a real process in the data. Classical inference requires a priori hypothesis. In this toy example, the hypothesis, \textit{i.e} the lack of separation of the two clusters, is based on clusters derived from the data. Moreover, here we force differences between groups of observations by clustering them, so the clustering results do not represent the true structure of the data. Thus, the discoveries are only the results of clustering algorithms and not those of a true biological signal due to this double use of the data and the bad structures forced by clustering. 
For example, in the context of RNA-seq data analysis, accounting for this clustering step during the inference step is one of the open problems in the eleven grand challenges in single-cell data science mentioned by \citet{lahnemann2020eleven}.

Our goal is to propose new methods for post-clustering inference that take into account the clustering step and the potential artificial differences it may introduce. For any clustering method that can be applied to all features of the data to build clusters, we are interested in testing the null hypothesis that a particular feature does not truly separate two of the estimated clusters. In particular, this null hypothesis allows that the feature: i) is not involved in the separation of the two subgroups and is not affected by the clustering step, and ii) is only involved in this separation because the clustering method applied to the data forced differences. 

Recently, some methodological work has been done on post-clustering inference. Since the data is used twice, many of them use selective inference \citep{tibshirani2016exact, lee2016exact} to account for the clustering step. Selective inference aims to control the selective type I error. This is defined as the probability under the null of rejecting the null hypothesis, given that the model and the null hypothesis have been selected thanks to the data. When data splitting is not possible, \citet{fithian2014optimal} has proposed to condition on this selection event during statistical hypothesis testing. In applying this approach, we use two different types of data: the data, to construct the model and the hypothesis, and the data given the fact that it has been used, i.e. data not yet observed, to perform the test. This leads to statistical hypothesis tests that efficiently control the selective type I error. Selective inference was first proposed for linear regression, change point detection \citep{jewell2019testing} and more recently for tree regression \citep{neufeld2021tree}. Clustering is also a framework in which selective inference has been applied recently. For post-clustering inference applied on RNA-seq data, \citet{zhang2019valid} have developed a truncated-normal statistic that use selective inference and leads to valid p-values under their null of no differential expression. However, in addition to selective inference, they use data splitting, which is only possible if the number of observations is large enough. They also use a supervised approach to predict the partition formed on half of the data on the remaining half. Instead of conditioning on the clustering event in their statistical hypothesis test, they condition on the fact that in the remaining half of the data, the labels of the observations are predicted thanks to a supervised approach. More recently, \citet{gao2020selective} have developed a multivariate selective test to investigate whether two estimated clusters are truly separated or whether the observations they contain come from a single cluster. By using selective inference, they account for the clustering step. Their approach is suitable for cluster validation because their null hypothesis is the equality of two cluster centers. This method also leads to valid p-values under the null hypothesis (Figure \ref{fig:1} \textbf{panel c}). However, this method is not suitable for our purpose, since in this particular context the goal is to study the separation of two clusters at the feature level, i.e., in a univariate setting. 

In this paper, we introduce three new methods for post-clustering inference. First, we adapt the method proposed by \citet{gao2020selective} for univariate hypotheses to investigate whether individual features contain information about group (clustering) structure. In doing so, we use a data-driven and fixed clustering of the data to ensure interpretations. To deal with the multiple clusters case, we also present an extension of this first test based on an aggregation of its p-values. Second, we propose another approach using a test of multimodality that account for the clustering step by investigating the presence of a continuum in the distribution of the variable. The paper proceeds as follows. In the \hyperref[sec:Methods]{Methods} section, we describe the methods we proposed for post-clustering inference. These approaches are then evaluated and compared in the \hyperref[sec:Res]{Results} section using extensive numerical simulations and a real ecological dataset. Some final comments can be found in the \hyperref[sec:Discussion]{Discussion} section.

\section{Methods}
\label{sec:Methods}
In the following, let $\matrice{X}$ be a $n \times p$ random variable of $n$ observations of $p$ features, with $g$\textsuperscript{th} column $\matrice{X}_g$. 
On $\matrice{X}$ we apply a clustering method $c()$ to create $c(\matrice{X})$, a partition of the $n$ observations into $K$ disjoint clusters $C_1, \dots, C_K$. We are interested in the ability of a given variable $\matrice{X}_g$ to separate two clusters $C_k$ and $C_l$ estimated using all the information contained in $\matrice{X}$ with the clustering method $c()$.

\subsection{Selective test}
To develop our statistical hypothesis testing, we first specify a generative model to the observations along $\matrice{X}_g$. We assume that each of the $n$ observations of $\matrice{X}_g$ comes from independent Gaussian distributions with unknown mean $\mu_{gi}$ and known variance $\sigma_g^2$. Then, for all $i \in \{1,\dots, n\}$, $X_{gi} \sim \mathcal{N}(\mu_{gi}, \sigma_g^2)$. Because of the independence between each $X_{gi}$, the multivariate distribution of $\matrice{X}_g$ is a multivariate Gaussian distribution $\mathcal{N}_n\left(\matrice{\mu_g}, \sigma^2_g\matrice{I_n}\right)$ with mean $\matrice{\mu_g} = \left(\mu_{g1}, \dots, \mu_{gn} \right)^t$ and covariance matrix $\matrice{\Sigma} = \sigma_g^2\matrice{I_n}$. 
Let $\matrice{x}_g$ be the realisation of $\matrice{X}_g$ observed in $\matrice{X}$. Now, for a cluster $C_k$, let $$\overline{\mu}_g^{C_k} = \frac{1}{|C_k|}\sum\limits_{i\in C_k} \mu_{gi} \qquad \mbox{and} \qquad \overline{X}_g^{C_k} = \frac{1}{|C_k|}\sum\limits_{i\in C_k} X_{gi}$$ 
be the true mean and empirical mean, respectively, of the variable $\matrice{X}_g$ in cluster $C_k$. Testing for a mean shift between the two clusters is a straightfoward way to evaluate the separation of two clusters along $\matrice{X}_g$. Thus, we define the two following hypotheses:
\begin{equation}
    \label{eq:test}
    \mathcal{H}_0 : \overline{\mu}_g^{C_k} = \overline{\mu}_g^{C_l} \qquad \mbox{vs} \qquad \mathcal{H}_1 : \overline{\mu}_g^{C_k} \neq  \overline{\mu}_g^{C_l}
\end{equation}

\noindent By introducing a contrast vector $\boldsymbol{\eta} \in \mathbb{R}^n$ defined by: $\eta_i = \frac{\mathbb{1}_{i \in C_k}}{|C_k|} - \frac{\mathbb{1}_{i \in C_l}}{|C_l|} \forall i =1,\dots, n$ following \citet{jewell2019testing, gao2020selective}, we can rewrite (\ref{eq:test}) above as: 
\begin{equation}
    \label{eq:test_2}
    \mathcal{H}_0 : \boldsymbol{\mu}_g^t\boldsymbol{\eta} = 0 \qquad \mbox{vs} \qquad \mathcal{H}_1: \boldsymbol{\mu}_g^t\boldsymbol{\eta} \neq 0
\end{equation}

$\mathcal{H}_0$ in (\ref{eq:test_2}) is actually generated by a function of the data $c(\matrice{X})$, which clearly sets us in the context of selective inference. Conditioning on this clustering event within statistical inference procedures is thus required. In particular, we derive an adaptation of the p-value proposed by \citet{jewell2019testing} (originally intended for change point detection) for our purposes of clustering:
\begin{equation}
    \label{eq:pvalue}
    p^{C_k, C_l}_g \equiv \mathbb{P}_{\mathcal{H}_0}\left(|\boldsymbol{X}_g^t \boldsymbol{\eta}| >|\boldsymbol{x}_g^t \boldsymbol{\eta}| \mbox{ | } C_k, C_l \in c(\boldsymbol{X})\right)
\end{equation}
Here we condition on the estimation of $C_k$ and $C_l$ by $c(\matrice{X})$, which leads to the definition of $\mathcal{H}_0$, and the resulting p-values (\ref{eq:pvalue}) account for the clustering as well as the uncertainty associated with the estimation of these two clusters. 
$p^{C_k, C_l}_g$ quantifies the probability that the mean difference between $C_k$ and $C_l$ is as large as the observed difference under $\mathcal{H}_0$ given the observed clustering structure. Its calculation relies on all possible realisations of $\matrice{X}_g$ resulting in the same estimation of $C_k$ and $C_l$ when we apply $c()$ to $\matrice{X}$. Yet, enumerating all such data sets $\matrice{X}$ is hard. To get more tractable p-values, we follow \citet{jewell2019testing} and \citet{gao2020selective} in constraining the randomness in the random variable $\matrice{X_g}$ and we define our p-value as follows:

\begin{equation}
    \label{eq:pvalue2}
    \tilde{p}^{C_k,C_l}_g \equiv \mathbb{P}_{\mathcal{H}_0}\left(|\matrice{X}_g^t\matrice{\eta}| >|\matrice{x}_g^t\matrice{\eta}| \mbox{ }| C_k, C_l \in c(\matrice{X}), \matrice{\pi_\eta}^\perp \matrice{X}_g = \matrice{\pi_\eta}^\perp \matrice{x}_g\right) 
\end{equation}
where $\boldsymbol{\pi}^\perp_{\boldsymbol{\eta}} = \boldsymbol{I}_n - \frac{\boldsymbol{\eta}\boldsymbol{\eta}^t}{\|\boldsymbol{\eta}\|^2_2}$ restricts the random variable $\matrice{X}_g$ to a space defined by the scalar $\matrice{\pi_\eta}^\perp \matrice{x}_g$ without losing control of type I error \citep{gao2020selective}. The p-value (\ref{eq:pvalue2}) can be rewritten as (see Supplementary Materials for the proof): 

\begin{equation}
    \label{eq:pvalue3}
    \tilde{p}_g^{C_k,C_l} = \mathbb{P}_{\mathcal{H}_0}\left(|\phi_g|>|\boldsymbol{x}_g^t\boldsymbol{\eta}| \mbox{ | }\phi_g \in S_g \right)
\end{equation}
where $S_g = \left\{\phi_g : C_k, C_l \in c\left(\boldsymbol{x}\left(\phi_g\right)\right)\right\}$ is the set of perturbations of the $g$\textsuperscript{th} variable from $\matrice{X}$ where both $C_k$ and $C_l$ are conserved by $c()$, and $\phi_g = \boldsymbol{X}_g^t\boldsymbol{\eta} \stackrel{\mathcal{H}_0}{\sim} \mathcal{N}\left(0, \sigma^2_g\|\boldsymbol{\eta}\|^2_2\right)$. $\boldsymbol{X}(\phi_g)$ thus represents a perturbed version of the data $\boldsymbol{X}$, where only the $g$\textsuperscript{th} variable is perturbed:
$$\boldsymbol{x_g} - \frac{\boldsymbol{\eta \eta}^t \matrice{x}_g}{\|\boldsymbol{\eta}\|_2^2}+ \frac{\boldsymbol{\eta} \phi_g}{\|\boldsymbol{\eta}\|_2^2}$$
This perturbation has a clear interpretation: if $|\phi_g| > |\matrice{x}_g^t\matrice{\eta}|$ data from the two clusters are split further apart along $\matrice{X}_g$ than is observed in the data; whereas if $|\phi_g| < |\matrice{x}_g^t\matrice{\eta}|$ instead, they are brought closer together along $\matrice{X}_g$ (and if $\phi_g = \matrice{x}_g^t\matrice{\eta}$ the data are actually not perturbed because in this case $\matrice{X}(\phi_g) = X$). Note that \eqref{eq:pvalue3} can be rewritten as $\mathbb{P}_{\mathcal{H}_0}\left(|\phi_g| > |\matrice{x}_g^t\matrice{\eta}|, \phi_g \in S_g\right)/{\mathbb{P}_{\mathcal{H}_0}\left(\phi_g \in S_g\right)}$. So if $C_k$ and $C_l$ can only be preserved when the observation are perturbed further apart, then (\ref{eq:pvalue3}) will be large since $\mathbb{P}_{\mathcal{H}_0}\left(|\phi_g| > |\matrice{x}_g^t\matrice{\eta}|, \phi_g \in S_g\right) \simeq \mathbb{P}_{\mathcal{H}_0}\left(\phi_g \in S_g\right)$. In conclusion, this selective test can be interpreted in terms of separability of the two clusters considered (even though it is based on a difference in means) as it boils down to quantifying the possibility to bring closer together the observations from the two clusters while preserving their separation.

In order to explicitly describe the set $S_g$ while retaining as much generality as possible about $c()$, we follow \citet{gao2020selective} and use Monte-Carlo simulations to approximate $\tilde{p}_g^{C_k,C_l}$. This strategy relies on \eqref{eq:pvalue3} being rewritten as:
\begin{equation}
    \label{eq:pvalueMC}
    \tilde{p}^{C_k,C_l}_g = \frac{\mathbb{E}\left[\mathbb{1}\left\{|\phi_g|>|\matrice{x}_g^t \matrice{\eta}|, \phi_g\in S_g\right\} \right]}{\mathbb{E}\left[\mathbb{1}\left\{\phi_g \in S_g\right\}\right]}
\end{equation}
Namely, we sample $\phi_g^1, \dots, \phi_g^N \overset{\text{i.i.d}}{\sim} \mathcal{N}\left(0, \sigma^2_g\|\matrice{\eta}\|_2^2\right)$ for some large value $N$, and replace the expectations in \eqref{eq:pvalueMC} with the sums over all samples. This Monte-Carlo procedure avoids the need to formally describe $S_g$. In order to enhance numerical efficiency, \citet{gao2020selective} use an importance sampling approach originally proposed by \citet{yang2016selective} to improve the likelihood of preserving the clustering in the perturbed data. Our proposed estimation of $\tilde{p}^{C_k,C_l}_g$ is thus:
\begin{equation}
    \label{eq:pvalueImpSamp}
        \tilde{p}_g^{C_k,C_l} \approx \frac{\sum\limits_{i=1}^N \pi^i \mathbb{1}\left\{|\omega_g^i| \geq |\matrice{x}_g^t\matrice{\eta}|, C_k, C_l \in c(\matrice{X}(\omega_g^i))\right\}}{\sum\limits_{i=1}^N \pi^i \mathbb{1}\left\{C_k, C_l \in c(\matrice{X}(\omega_g ^i))\right\}}
\end{equation}
 where $\omega_g^1, \dots, \omega_g^N \sim \mathcal{N}\left(\matrice{x}_g^t\matrice{\eta}, \sigma^2_g\|\matrice{\eta}\|^2_2\right)$, and $\pi^i = \frac{f_1(\omega_g^i)}{f_2(\omega_g^i)}$ the importance sampling probabilities with $f_1$ the density of a $\mathcal{N}\left(0,\sigma^2_g\|\matrice{\eta}\|^2_2\right)$ distribution and $f_2$ the distribution of a $\mathcal{N}\left(\matrice{x}_g^t\matrice{\eta},\sigma^2_g\|\matrice{\eta}\|^2_2\right)$ distribution. Of note, we adapt the method from \citet{phipson2010permutation} to obtain unbiased Monte-Carlo p-values estimations (see Supplementary Materials for details).

At the core of the above test is the scaling variance parameter $\sigma^2_g$, which represents the variance of each column of $\matrice{X}_g$. While $\sigma^2_g$ is assumed to be known in the test, it is not the case in practice and we propose to use the following plug-in estimate instead: 
$$\hat{\sigma}^2_g = \frac{1}{|C_k| + |C_l| - 1}\sum\limits_{i\in C_k, C_l}\left(X_{gi} - \overline{X}_g^{C_k, C_l}\right)^2 \qquad \text{with} \quad \overline{X}_g^{C_k, C_l} = \frac{1}{|C_k| + |C_l|} \sum\limits_{i \in C_k, C_l}X_{gi}$$
This variance estimate only takes into account observations from the two clusters of interest in the test, in line our null hypothesis of no separation of the two clusters (the variance itself can informs on the separation of the data \citep{liu2010understanding}). In some cases, this $\hat{\sigma}^2_g$ could underestimates the variance of $\matrice{X}_g$ (particularly if the clustering induces strong artificial differences). Meanwhile, \citet{gao2020selective} rely on a different estimate, that instead overestimate the variance in certain cases. Still they have showed type I error control is guaranteed, even with an overestimated variance, at the cost of being overly conservative (see Supplementary Figure S1 for additional details). 

\subsection{A more powerful test in the presence of intervening clusters}

The above selective test has been designed for comparing a pair of clusters. Yet, in practice there are often more than $2$ clusters. In such case, the test could have very limited statistical power even for well separated clusters: if there is one or more additional cluster in-between the two clusters of interest, it quickly becomes impossible to perturbed them closer together without changing the clustering (see Supplementary Figure S2). To overcome this limitation, we extend the above selective test assuming that two estimated clusters $C_k$ and $C_l$ are separated on $\matrice{X}_g$ if and only if at least one of the adjacent cluster pairs in-between them are separated. This means that on the contrary, if there is a continuum on $\matrice{X}_g$ to go from $C_k$ to $C_l$, then $C_k$ to $C_l$ are not separated. By testing only the separation of pairs of adjacent clusters we retain the statistical power of the selective test. We propose to combine all the in-between adjacent pair selective test p-values into a so called combined selective test to finally assess the separation of $C_k$ and $C_l$ on $\matrice{X}_g$. 

To identify the clusters in-between $C_{k}$ and $C_{l}$ on $\matrice{X}_g$, we define the set:
$$ \mathcal{C}^{k:l}_g := \left\{ C_i, i = 1, \dots, K ~/~ \overline{X}_g^{C_i} \in \left[ \min\left(\overline{X}_g^{C_{k}}, \overline{X}_g^{C_{l}}\right), \max\left(\overline{X}_g^{C_{k}}, \overline{X}_g^{C_{l}}\right)\right]\right\}$$
where $\overline{X}_g^{C_i} = \frac{1}{|C_i|}\sum\limits_{j \in C_i}X_{gj}$, implicitly sorting clusters according to their empirical mean on $\matrice{X}_g$.
We define two clusters $C_{m_{1}}$ and $C_{m_{2}}$ as adjacent on $\matrice{X}_g$ if $\mathcal{C}_g^{m_1:m_2} = \left\{C_{m_{1}}, C_{m_{2}}\right\}$. So if $|\mathcal{C}_g^{k:l}| = M$, there are $M-1$ pairs of adjacent clusters in $\mathcal{C}_g^{k:l}$ that draw a path from $C_k$ to $C_l$. We can then define: 
$$p_g^{C_k:C_l} := f\left(p_g^1, \dots, p_g^{M-1}\right)$$
where $f$ must be a merging function as described in \citet{vovk2020combining}.
Based on numerical simulations, we favor the use of the harmonic mean merging function,  recommended by \citet{vovk2020combining} for potential dependencies between the p-values \--- which is our case here since each cluster data contributes to two p-values \--- and features a good trade-off between type I error and statistical power (see Supplementary Figure S3). Thus, we use:
$$p_g^{C_k:C_l} = \min\left(e\log{(M-1)}\frac{M-1}{\sum\limits_{i=1}^{M-1}\frac{1}{p_g^i}}, 1\right)$$ 
Of note, in order for all $p^1_g,\dots,p^{M-1}_g$ to be computed using the same variance estimate, we propose this time to plug-in an estimate of $\sigma^2_g$ that accounts for all observations belonging to either one the adjacent clusters in $\mathcal{C}^{g:l}_g$: $$\hat{\sigma}_g^2 = \frac{1}{|C_g^{k:l}|-1} \sum\limits_{C\in C_g^{k:l}}\sum\limits_{i\in C}\left(X_{gi} - \overline{X}_g^{C_g^{k:l}}\right)^2 \qquad \text{with} \quad \overline{X}_g^{C_g^{k:l}} = \frac{1}{|C_g^{k:l}|} \sum\limits_{C \in C_g^{k:l}}\sum\limits_{i\in C}X_{gi}$$

\subsection{Multimodality test}

The separation of two clusters according to a given variable is equivalent to this variable's distribution being multimodal. Following \citet{kim2021marcopolo}, multimodality thus becomes a marker for the separation of clusters: each mode corresponds to a group of homogeneous observations (i.e., a cluster), separated by less dense regions of the distribution. But as with artificial mean differences arising from clustering, multimodality may also be an artefact caused by the clustering method $c()$. We propose to leverage this notion of continuum between two clusters: if $C_k$ and $C_l$ are separated, then there must be dip in the distribution of this variable at some point between the two (i.e. multimodality). On the other hand, if there is a continuum between these two clusters, then they cannot truly be separated (i.e. unimodality). Fortunately, such a continuum cannot be caused by the clustering method.

This second proposal can be seen as a simplified version of our first selective test. Indeed, by perturbing the data in the selective test to see if we can bring the two clusters closer without changing the clustering, we assess how likely it would be to observe a continuum between $C_k$ and $C_l$. If there is a continuum between $C_k$ and $C_l$ on $\matrice{X}_g$, then its distribution must be unimodal. Thus, to investigate separability of those two clusters on $\matrice{X}_g$, it suffices to apply a unimodality test to its distribution restricted only to the individuals from clusters of the set $C_g^{k:l}$. Indeed, if the $\matrice{X}_g$ separates $C_{k}$ and $C_{l}$, then there are at least two clusters in $C^{k:l}_g$ that are separated from each other, and in particular, since these clusters are between $C_{k}$ and $C_{l}$, there is also a separation between them on $\boldsymbol{X}_g$.

A unimodality test compares the null hypothesis ``distribution of $\matrice{X}_g$ is unimodal'' to the alternative ``distribution of $\matrice{X}_g$ is multimodal''. In the context of unsupervised clustering, \citet{kalogeratos2012dip} developed a clustering algorithm based on incremental unimodality testing, and \citet{siffer2018your} developed unimodality a test to assess data clusterability based on their multivariate distribution. \citet{ameijeiras2021multimode} give a recent overview on unimodality testing, but three tests are the most frequent: i) the Silverman test \citep{silverman1981using} based on the kernel estimate of the density $f$ of the data,  ii) the Dip Test \citep{hartigan1985dip} based on the cumulative distribution function $F$, and iii) the excess mass test \citep{muller1991excess}. The Dip Test avoids the need for estimating of additional parameters or making any distributional assumption and has already been applied to clustering \citep{kalogeratos2012dip, wasserman2014feature, schelling2020dataset}. Furthermore, compared to several multimodality tests available in the R package \texttt{multimode} \cite{ameijeiras2021multimode}, the Dip Test outperforms its competition both in terms of computation times and performances (see Supplementary Figure S4).


The Dip Test from \citet{hartigan1985dip} relies on the dip statistic $\displaystyle \operatorname{dip}(F) = \min_{G \in \mathcal{U}}\rho(F,G)$,
where $\rho(F,G) = \sup_x|F(x) - G(x)|$ and $\mathcal{U}$ is the class of unimodal distributions. Thus, the dip statistic is the distance of $F$ to the set of unimodal functions and it measures the deviation of our distribution from unimodality. If $F$ is unimodal, then $\operatorname{dip}(F) = 0$, and conversly if $F$ is multimodal, then $\operatorname{dip}(F) > 0$. In practice p-values can be computed as: 
$$p_{\widehat{D}_{n}}:= \mathbb{P}\left(d_{U_n} \geq \widehat{D}_{n}\right)$$ where $d_{U_n}$ is the dip statistic computed for a $n$-sample drawn from $\mathcal{U}[0,1]$ (the standard uniform distribution), $\widehat{D}_n$ is the observed dip statistic, and $n$ is the sample size. \citet{hartigan1985dip} showed that the Uniform distribution is the unimodal distribution with the asymptotically largest dip statistic among the unimodal distributions (intuitively the least favourable candidate for unimodality): a distribution with a dip statistic larger than that of the uniform distribution cannot be unimodal. Thus
$p_{\widehat{D}_{n}}$ is interpreted as the probability under the null case of unimodality that the uniform distribution has a dip statistic greater than the observed dip statistic of $\widehat{F}_n$. 

For our purposes, we apply the Dip Test to the distribution of the variable $\matrice{X}_g$ restricted to the individuals that are in the clusters of the set $C_{k:l}$ to test for a continuum between $C_{k}$ and $C_{l}$.

\section{Results}
\label{sec:Res}

\subsection{Numerical simulations study}

We present here results evaluating the behaviour of our proposed tests in the \hyperref[sec:Methods]{Methods} section both in terms of type-I error control and statistical power. 


\subsubsection{Behaviour in a two-dimensional setting}

\begin{figure}[htbp]
    \centering
    \includegraphics[width = .75\textwidth]{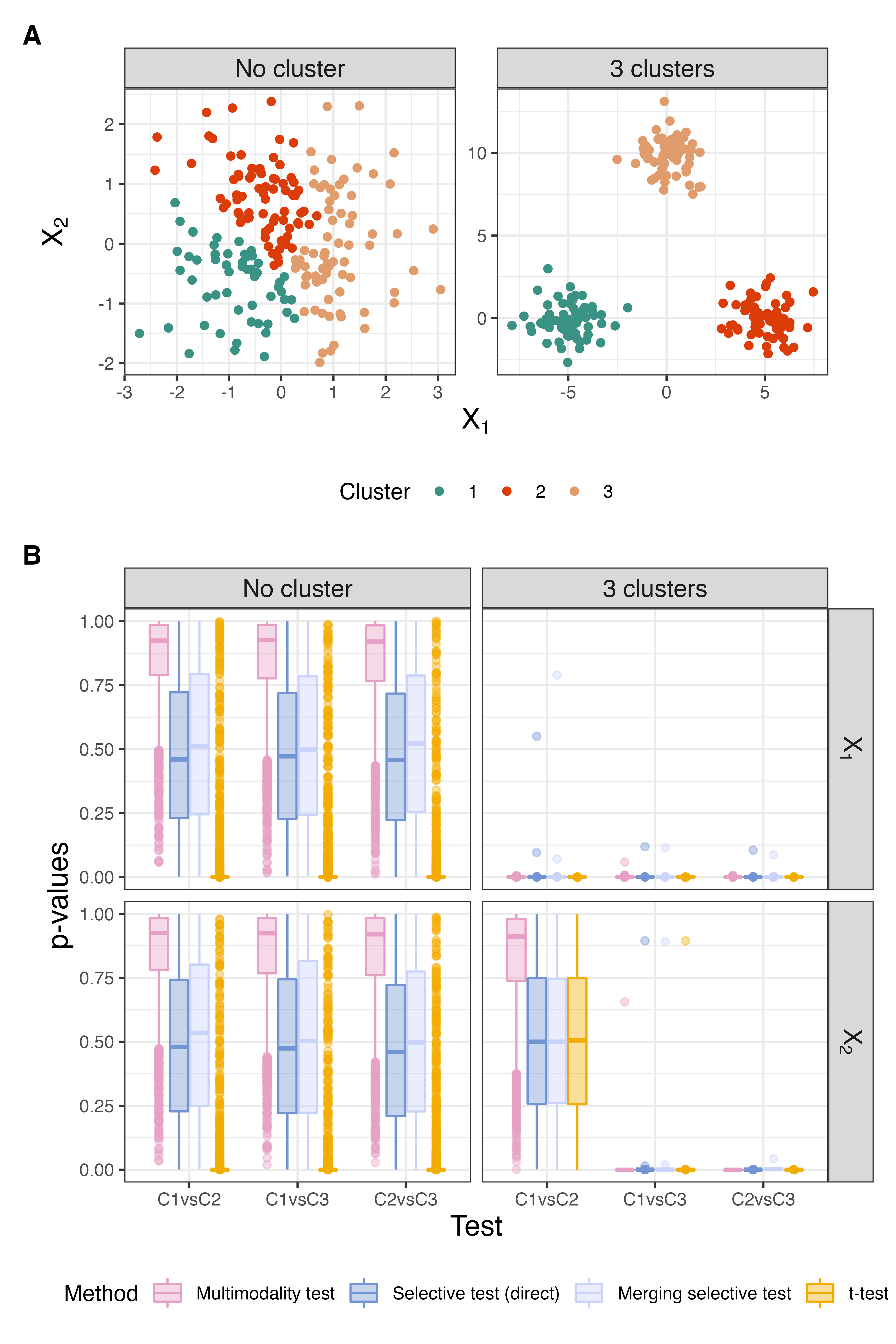}
    \caption{Validity of p-values returned by our proposed tests, comparison with t-test. \textbf{panel A} Data generation process. Two cases are studied: a case under a global null of no clusters in the data (No cluster) and a case with three real clusters (3 clusters). In both cases, hierarchical clustering with Ward method and Euclidean distance is used to build $3$ clusters.  \textbf{panel B} Resulting p-values for each possible cluster pair for each variable for $2000$ simulations of the data.}
    \label{fig:simuvalidity}
\end{figure}

We generated two-dimensional data ($p=2$) under two scenarios: (i) first with no separated clusters from a common standard Gaussian distribution $\mathcal{N}(0,1)$; and (ii) second with three clusters from Gaussian distributions $\mathcal{N}(\mu^{C_j}, 1)$ and built-in mean differences $\mu^{C_1} = (-5, 0)$, $\mu^{C_2} = (5, 0)$, and $\mu^{C_3} = (0, 10)$ (thus $X_1$ separated all three clusters while $X_2$ only separated $C_3$ from the rest, meaning $X_2$ was under the null when comparing $C_1$ and $C_2$). In both cases, we applied hierarchical clustering with Ward method and Euclidean distance to build three clusters. Figure \ref{fig:simuvalidity}A shows an example realisation for each scenario. In the first scenario, clusters were estimated by forcing differences between groups of observations, creating artificial differences between clusters, while in the second scenario, the estimated clusters represented the true structure of the data (each cluster weas a homogeneous and separate group of observations).

Figure \ref{fig:simuvalidity}B shows the results of the three proposed approaches compared to the p-values from the usual t-test for 2,000 repeated simulations each with a sample size of $n=200$. For the no cluster scenario, the t-test yielded extremely small p-values which translates into a direct inflation of the type-I error. In fact, the t-test identified the artificial differences created during the clustering process. Taking into account this clustering step, the p-values of the selective test $p_g^{C_k,C_l}$ and the p-values resulting from its merging extension $p_g^{C_k:C_l}$ were fairly uniformly distributed over $[0,1]$, ensuring a good calibration of the p-values and a control of type-I error. As for the multimodality test, its p-values were overly conservatives but consistent with no real separation of clusters. This was due to its reference being the Uniform distribution (the limit case for a unimodal distribution) while the data were generated from a Gaussian distribution (which has a lower dip statistic than the uniform distribution). Those good results were confirmed under the 3 clusters scenario when comparing $C_1$ and $C_2$ along $X_2$. For all other comparisons under this scenario, all 4 tests correctly detected the separated clusters that are under $\mathcal{H}_1$.


Of note, if clustering does not artificially force differences between groups of observations, e.g by discovering the actual group structure in the data, the t-test also control the type-I error. This illustrates the connections between artificial differences and the estimation of the number of clusters. But, this process is still testing data driven hypothesis, which does not respect the classical inference setting where hypothesis must be specified without using the data.

\subsubsection{Statistical power}

\begin{figure}[htbp]
    \centering
    \includegraphics[width = .75\textwidth]{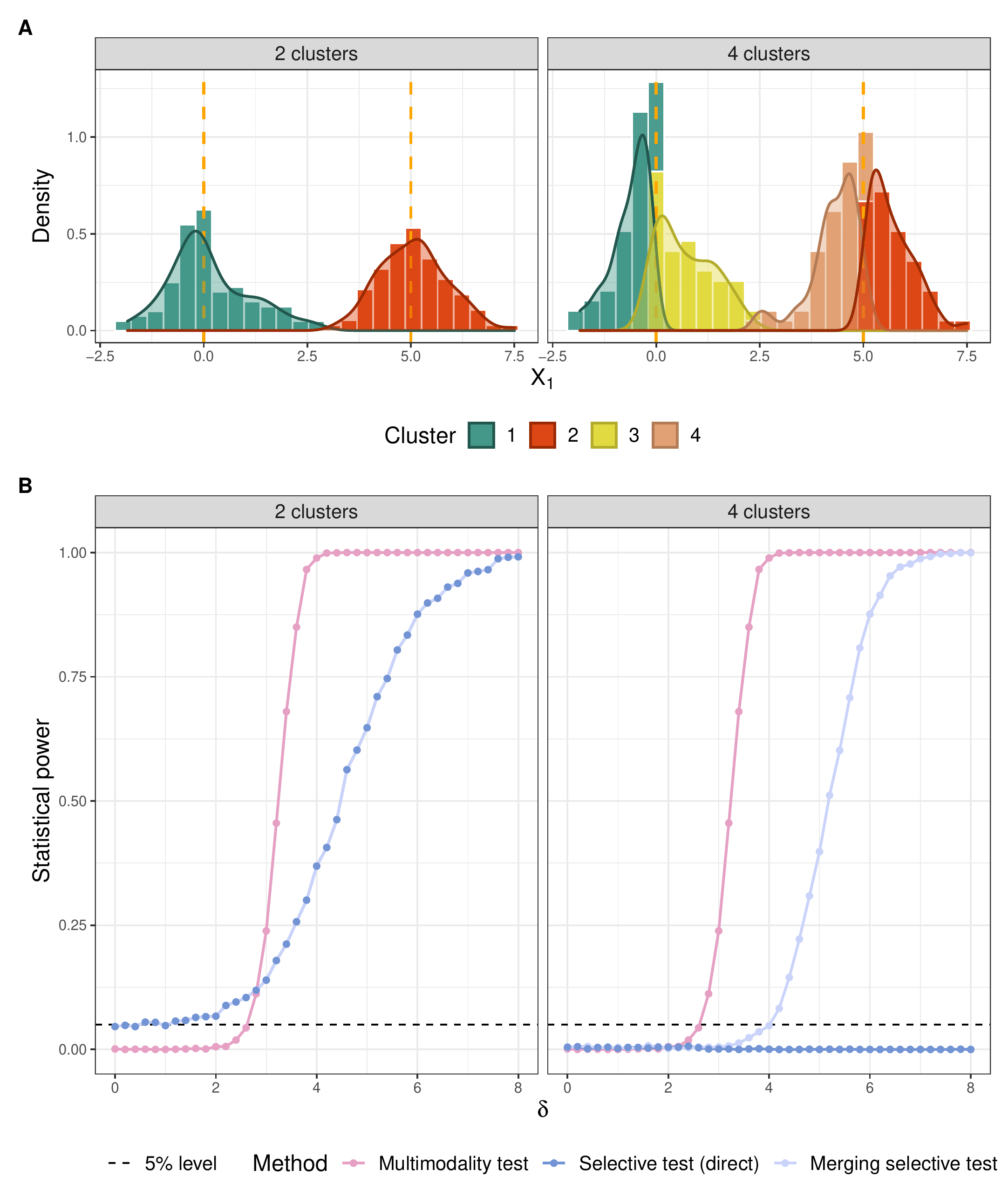}
    \caption{Statistical power of the proposed tests.
    \textbf{panel A} Data generation process: data are generated according to a univariate gaussian mixture with two components (with equal probability and variance) separated by a mean difference $\delta$ (contamination model). Two cases are studied: a case where the true numbers of clusters is estimated ($2$ clusters) and a case where more clusters are estimated ($4$ clusters). The orange dashed line represents the mean of the component. 
    \textbf{panel B} Statistical power ($5\%$ level) of the proposed tests for the separation of Cluster 1 and Cluster 2 according the mean difference $\delta$ separating the two components of the gaussian mixture.} 
    \label{fig:simu_power}
\end{figure}

We now generated data from a univariate mixture of two Gaussian distributions with equal proportions and variance: $0.5 \mathcal{N}(0,1) + 0.5 \mathcal{N}(\delta, 1)$, where the two components were separated by a mean difference of $\delta$ \--- also called a contamination model \citep{laurent2018multidimensional}. The two components distinguished two different clusters, and the magnitude of $\delta$ tuned their separability. We applied hierarchical clustering with Ward method and Euclidean distance to build either $2$ or $4$ clusters. Figure \ref{fig:simu_power}A displays an example realization of this simulation. In the 2-cluster case, the true structure of the data was uncovered, while in the 4-cluster case spurious clusters were introduced.
We evaluated the statistical power of the three proposed tests to detect the separation between clusters $1$ and $2$, the two most extreme clusters in the distribution of the data, according to $\delta$ at significance levels $\alpha = 5\%$ using $N = 2000$ Monte-Carlo replicates. 

Figure \ref{fig:simu_power}B displays the results. Intuitively, statistical power increases with $\delta$. The multimodality test appeared the most powerful in this setting , especially when $\delta \geq 3$. \citet{siffer2018your} has shown that $\delta = 3.04$ is a threshold value above  which bimodality begins to appear in such two-part Gaussian mixture. Moreover, since all clusters in the $4$-cluster case were between cluster $1$ and cluster $2$, statistical power achieved by the multimodality test is exactly the same as in the $2$-clusters case. In the two-clusters case, the direct selective test and the merging selective test had the same statistical power (since only two clusters were estimated, they were necessarily adjacent and therefore the direct selective test was exactly the same as the merging one).
Meanwhile, the direct selective test failed in the $4$-cluster case, regardless of the value of $\delta$. Indeed, it was impossible to bring clusters $1$ and $2$ closer without mixing them with clusters $3$ and $4$. Fortunately, the merging selective test avoided this pitfall, because the direct selective inference test performed favourably on adjacent clusters, and carried over the separation between clusters 3 and 4. Of note, the merging selective test remains valid when the numbers of adjacent p-values increases.

\subsection{Application to real ecological data}

To further assess our proposed approaches, we also analyzed real data available from the \texttt{R} package \texttt{palmerpenguins} \citep{horst2020penguins}. This benchmark dataset features $p=4$ measured variables \--- bill length (mm), bill depth (mm), flipper length (mm), and body mass (g) \--- for $344$ penguins. After removing observations containing missing values for at least one of the $4$ variables, $n=333$ observations were kept in our analysis. The penguins belonged three different species: Adelie, Chinstrap, and Gentoo (with $146$, $68$ and $119$ observations respectively). 

\subsubsection{Negative control}

We initially selected only female Gentoo penguins to create a negative control dataset. Since this dataset contained only observations of the same species and sex, there should be no real differences between observations. We applied hierarchical clustering using Ward method and Euclidean distance on scaled data to build $3$ clusters (scaling avoids the variable with the largest variance to dominate the clustering). Since there was no information defining any group structure in this subset, the clustering artificially created differences. Table \ref{table:negctrl} presents the p-values of each of the 3 proposed test along with the ones from the t-test for all cluster pairs along each of the four measures. Once again, the t-test identified numerous spurious associations. Meanwhile, all 3 proposed tests behaved properly by not identifying any of the four measures to be significantly separating clusters.
 
\begin{table}[!htbp]
\footnotesize
\begin{center}
\begin{tabular}{lcccc}
\hline
 \textbf{Cluster pair tested} & \textbf{Selective test (direct)} & \textbf{Merging selective test} & \textbf{Multimodality test} & \textbf{t-test} \\ 
    \quad Variable tested &  &  &  
    \\\hline

\textbf{Cluster 1 vs Cluster 2}\\
\quad bill length    & 0.4082                                                                                                       & 0.4110                                                                                                        & 0.4899                                          & 0.0759\phantom{*}          \\
\quad bill depth     & 0.6478                                                                                                       & 0.6400                                                                                                        & 0.1478                                          & 0.4802\phantom{*}          \\
\quad flipper length & 0.1160                                                                                                       & 0.1154                                                                                                        & 0.0992                                          & 0.0017*         \\
\quad body mass\smallskip       & 0.3321                                                                                                       & 0.3425                                                                                                        & 0.8320                                          & 0.0000*\\ 
\textbf{Cluster 1 vs Cluster 3}                                                                                                                                                                                                                                                                             \\
\quad bill length    & 0.1748                                                                                                       & 0.4995                                                                                                        & 0.6345                                          & 0.0001*         \\
\quad bill depth     & 0.2914                                                                                                       & 0.3025                                                                                                        & 0.5242                                          & 0.0000*         \\
\quad flipper length & 0.3361                                                                                                       & 0.3206                                                                                                        & 0.6146                                          & 0.0005*         \\
\quad body mass\smallskip       & 0.3404                                                                                                       & 0.3868                                                                                                        & 0.2918                                          & 0.1190\phantom{*}\\
\textbf{Cluster 2 vs Cluster 3}                                                                                                                                                                                                                                                                             \\
\quad bill length    & 0.2096                                                                                                       & 0.2120                                                                                                        & 0.9140                                          & 0.0041*         \\
\quad bill depth     & 0.1867                                                                                                       & 0.6618                                                                                                        & 0.2376                                          & 0.0000*         \\
\quad flipper length & 0.2101                                                                                                       & 0.4322                                                                                                        & 0.1337                                          & 0.0000*         \\
\quad body mass       & 0.1573                                                                                                       & 0.7967                                                                                                        & 0.6759                                          & 0.0000*         \\ \hline
\end{tabular}
\end{center}
\caption{\textbf{P-values for all cluster pair tests along each of the 4 variables from the negative control real data}. \\
\textrm{*} highlights significant p-values at the $\alpha = 5 \%$ level}
\label{table:negctrl}
\end{table}

\subsubsection {Full data analysis}
\begin{figure}[!htbp]
    \centering
\includegraphics[width = .9\textwidth]{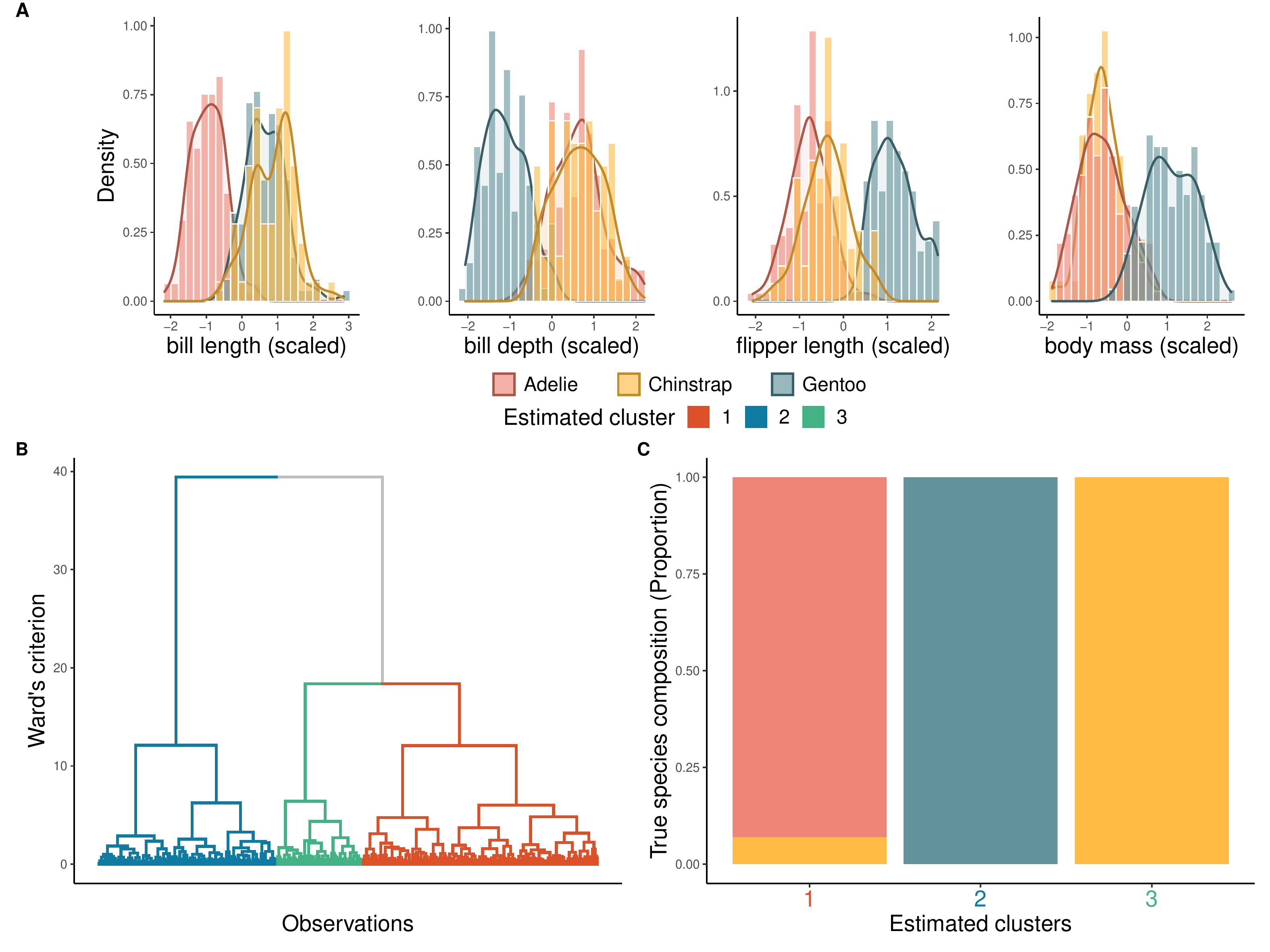}
    \caption{Clustering of the 333 completely observed palmer penguins.
    \textbf{A} Distribution of each scaled variable according to the three true species of penguins.
    \textbf{B} $3$ clusters are build thanks to hierarchical clustering. Colors in the dendogram represent the estimated clusters. 
    \textbf{C} Clustering results reveals that estimated clusters correspond to true species}
    \label{fig:fig_appli}
\end{figure}

We now included all $n=333$ penguins in our analysis and analyzed the data as if we didn't know the species of the penguins. Since all 3 species were present, we want to identify which features are actually separating them. Figure \ref{fig:fig_appli}A displays the density distribution for the four (scaled) variables across all 3 species. Adelie and Chinstrap penguins appear harder to distinguish, as they only appear to differ in bill length: Chinstrap penguins have larger bills (comparable to those of Gentoo penguins) than Adelie penguins. The Gentoo penguin species is the easiest to identify, as it is clearly different in the other 3 measures. Once again, we applied hierarchical clustering to scaled data with Euclidean distance and Ward method. Figure \ref{fig:fig_appli}B displays the results of this clustering where we cut the dendrogram to get three clusters. Those three estimated clusters recovered the true species: cluster $2$ and cluster $3$ each contained only penguins of the Gentoo and Chinstrap species (respectively) while cluster $1$ contained a mixture of two species ($100 \%$ of the Adelie penguins plus $11$ Chinstrap penguins). 

Table \ref{tab:allpeng} presents the p-values of each of the 3 proposed test along with the ones from the t-test for all cluster pairs along each of the four measures. Since the identified clusters corresponded to the three real penguin species, the clustering step was not expected to induce any artificial differences, and thus the t-test results can be used as reference. 
Only one comparison was not significant at the $5\%$ level according to the t-test: bill depth did not separate Adelie (cluster $1$) from Chinstrap penguins (cluster $3$), which was visually coherent with Figure \ref{fig:fig_appli}A.
The multimodality test seemed to lack statistical power here, but inspection of the measures distribution depicted in Figure \ref{fig:fig_appli}A showed that only a few comparison exhibited multimodality (namely cluster 1/Adelie compared to either two other clusters along flipper length. Both selective tests identified more significant differences (6/11 for the direct test and 7/11 for the merging test which is more robust when more than 2 clusters are identified). The missed separations can be explained by the lack of statistical power to detect small difference (see Supplementary Table 1 for additional details). By accounting the clustering step, our proposed test could have a reduced power compared to other tests like t-test (which is is the uniformly most powerful test \citep{lehmann2012some}). But, this is because they are appropriately accounting for the variability and the uncertainties of the clustering step leading to results that are always valid.  

\begin{table}[!hbtp]
\footnotesize
\begin{center}
\begin{tabular}{lcccc}
\hline
 \textbf{Cluster pair tested} & \textbf{Direct selective test} & \textbf{Merging selective test} & \textbf{Multimodality test} & \textbf{t-test} \\ 
    \quad Variable tested &  &  &  
    \\\hline
    \textbf{Cluster 1 vs Cluster 2} \\ 
    \textbf{\textit{(Adelie vs Gentoo)}}\\ 
\quad bill length    & 0.0024*                                                                                                      & 0.0023*                                                                                                       & 0.1647\phantom{*}                                          & 0*\phantom{.0000}               \\
\quad bill depth    & 0.0015*                                                                                                      & 0.0017*                                                                                                       & 0.3687\phantom{*}                                          & 0*\phantom{.0000}               \\
\quad flipper length & 0.0725\phantom{*}                                                                                                       & 0.1832\phantom{*}                                                                                                        & 0.0047*                                         & 0*\phantom{.0000}               \\
\quad body mass\smallskip       & 0.0439*                                                                                                      & 0.0008*                                                                                                       & 0.6402\phantom{*}                                          & 0*\phantom{.0000}               \\ 
\textbf{Cluster 1 vs Cluster 3}\\
\textbf{\textit{(Adelie vs Chinstrap)}} \\
\qquad bill length    & 0.1748                                                                                                       & 0.0191*                                                                                                       & 0.0674\phantom{*}                                          & 0*\phantom{.0000}              \\
\qquad bill depth      & 0.2266                                                                                                       & 0.2323\phantom{*}                                                                                                        & 0.2373\phantom{*}                                          & 0.0702\phantom{*}          \\
\qquad flipper length & 0.4318                                                                                                       & 0.4434\phantom{*}                                                                                                        & 0.0168*                                         & 0*\phantom{.0000}              \\
\qquad body mass\smallskip       & 0.7036                                                                                                       & 0.7027\phantom{*}                                                                                                        & 0.3311\phantom{*}                                          & 0.0267*        \\ 
\textbf{Cluster 2 vs Cluster 3}\\
\textbf{\textit{(Gentoo vs Chinstrap)}}\\ 
\qquad bill length   & 0.2263\phantom{*}                                                                                                       & 0.2115\phantom{*}                                                                                                        & 0.0927                                          & 0*\phantom{.0000}               \\
\qquad bill depth     & 0.0084*                                                                                                      & 0.0051*                                                                                                       & 0.2245                                          & 0*\phantom{.0000}               \\
\qquad flipper length & 0.0186*                                                                                                      & 0.0205*                                                                                                       & 0.1585                                          & 0*\phantom{.0000}               \\
\qquad body mass     & 0.0002*                                                                                                      & 0.0002*                                                                                                       & 0.4174                                          & 0*\phantom{.0000}               \\ \hline
\end{tabular}
\end{center}
\caption{\textbf{P-values for all cluster pair tests along each of the 4 variables from the positive control real data}. \\
\textrm{*} highlights significant p-values at the $\alpha = 5 \%$ level}
\label{tab:allpeng}
\end{table}

\section{Discussion}
\label{sec:Discussion}

In this paper, we propose three new statistical tests for post-clustering inference that can be used to identify variables that separate two estimated clusters. We show that the double use of data (for clustering and for inference) and the failure to propagate the uncertainty associated with cluster estimation can lead to invalid p-values. This is particularly the case when too many clusters are estimated compared to the true underlying structure of the data. In this case, since there is no true process separating every estimated clusters, the clustering forces artificial differences between observations belonging to a common group of observations. Our three proposed tests, which take into account this clustering step and/or its possible impact on inference, give p-values that indicate a separation not induced by the clustering algorithm but emanating from the underlying data generating process, while controlling the Type-I error rate adequately. Our approaches can be used regardless of the chosen clustering algorithm and take into account many data analysis pipelines where clustering results are used post-hoc to describe and interpret clusters. 


All three approaches test whether there is a separation between clusters along a given variable. The selective test is a rigorously defined test based on the concepts of selective inference, adapted from the seminal work of \citet{gao2020selective} and makes a Gaussian assumption on the data. Although it tests a univariate mean difference between two clusters, it also exploits the multivariate structure of the data since the (perturbed) clustering uses all variables. The multimodality test, on the other hand, is based on the more intuitive concept of multimodality to characterise the separation of two clusters along a variable. It only relies on univariate considerations, as the separation of clusters is examined based on the distribution of each variable. Thus, unlike the selective test, which has longer computation times (dependent on the number of observations, the number of variables, and the number of Monte-Carlo simulations required to estimate p-values), it is very computationally efficient (see Supplementary Figure S5). However, this simplicity comes at the expense of a larger null hypothesis: the multimodality test requires a clear separation between clusters on the variable to work well, as it only uses the variable-level information and does not consider the entire structure of the data. Finally, since false-negative problems could occur with the selective test (particularly when the two clusters of interest are separated by other clusters), we also propose a merging method based on the aggregation of p-values. This method has the advantage of correcting these false positive problems while guaranteeing good statistical power. However, its computation cost is even greater than the selective test because this approach requires the computation of all the adjacent p-values between $C_k$ and $C_l$.  


The selective test rely on some distributional assumptions. In particular, because it uses the selective inference framework, it assumes Gaussian data to efficiently control the Type I error rate. We show that the selective test remains robust to other distributions (see Supplementary Figure S6). The multimodality test is based on the Dip Test which is a non-parametric test of unimodality. However, in practice, its p-value is computed using the Uniform distribution as the reference distribution under the null of unimodality. It could affect its statistical power but the control of the Type I error is still guaranteed. 

The main limitation of our tests lies in the high dimensional setting. Due to the large number of variables and their correlation, perturbation-based approaches can fail. In our case, this result is amplified by the fact that the perturbation is only univariate. Thus, the selective inference test performs poorly in high dimension since it is exclusively based on perturbations. In addition, the calculation of the p-values is done using a Monte-Carlo approach requiring the clustering step to be repeated for each simulation and for each variable. So, if the number of variables is high, the computation time of the selective test can therefore be too long. Another problem of our approaches is related to the "signal vs. noise" ratio of the high dimension. In high dimension, a small signal repeated over a large number of variables is sufficient to create separated clusters in the high dimensional space \citep{klawonn2012clusters}. For example in a two-components gaussian mixture, for $n=100$ observations, a mean difference $\delta = 1$ repeated over $p=50$ variables is enough to generate separated clusters on the first principal component of a PCA. However, the unimodality test is not powerful enough when the signal is too weak. The problem here is that the existence of clusters is only due to the repetition of the signal on a large number of variables, \textit{i.e.} one has to take into account all the variables and the information they bring to explain the separation between clusters, but the unimodality test is purely univariate, being only interested in the information brought by the tested variable and this is why it lacks power in high dimension. Therefore, all the issues raised by the high-dimension constitute a natural path to apply and extend the results of our work presented here.

\newpage
\bibliographystyle{unsrtnat}
\bibliography{references.bib}

\appendix

\newpage

\section{Proof: Computation of the selective p-values}
\label{sec:proof}
We  want to compute the selective p-value given in \eqref{eq:pvalue2}: 

\begin{equation*}
    p_g^{C_k,C_l} \equiv \mathbb{P}_{H_0}\left(|{\boldsymbol{X}_g}^t\boldsymbol{\eta}| >|{\boldsymbol{x}_g}^t\boldsymbol{\eta}| \mbox{ }| C_k, C_l \in c(\boldsymbol{X}), \boldsymbol{\Pi_\eta}^\perp \boldsymbol{X}_g = \boldsymbol{\Pi_\eta}^\perp \boldsymbol{x}_g\right) 
\end{equation*}
To compute (\ref{eq:pvalue2}), we have to write our data matrix $\boldsymbol{X}$ as a function of our statistic ${\boldsymbol{X}_g}^t\boldsymbol{\eta}$ and the residual term $ \boldsymbol{\Pi_\eta}^\perp \boldsymbol{X}_g$ where $\boldsymbol{\Pi_\eta}^\perp = \boldsymbol{I}_n- \frac{\boldsymbol{\eta}\boldsymbol{\eta}^t}{\|\boldsymbol{\eta}\|_2^2}$
Since, $\boldsymbol{X}_g = \boldsymbol{\Pi_\eta}^\perp\boldsymbol{X}_g + \left(\boldsymbol{I}_n-\boldsymbol{\Pi_\eta}^\perp\right)\boldsymbol{X}_g$, then :
\begin{align*}
    c(\boldsymbol{X}) &= c\left(\left[\boldsymbol{x}_1|\dots|\boldsymbol{X}_g|\dots|\boldsymbol{x}_p\right]\right) \\
    &= c\left(\left[\boldsymbol{x}_1|\dots|\boldsymbol{0}_n|\dots|\boldsymbol{x}_p\right] + \left[\boldsymbol{0}_n|\dots|\boldsymbol{X}_g|\dots|\boldsymbol{0}_n\right] \right) \\
    &= c\left(\left[\boldsymbol{x}_1|\dots|\boldsymbol{0}_n|\dots|\boldsymbol{x}_p\right] + \left[\boldsymbol{0}_n|\dots|\boldsymbol{\Pi}^\perp_{\boldsymbol{\eta}} \boldsymbol{X}_g + (\boldsymbol{I}_n - \boldsymbol{\Pi_\eta}^\perp )\boldsymbol{X}_g|\dots|\boldsymbol{0}_n\right] \right) \\
    &= c\left(\left[\boldsymbol{x}_1|\dots|\boldsymbol{0}_n|\dots|\boldsymbol{x}_p\right] + \left[\boldsymbol{0}_n|\dots|\boldsymbol{\Pi}^\perp_{\boldsymbol{\eta}} \boldsymbol{X}_g + \left(\boldsymbol{I}_n - \boldsymbol{I}_n + \frac{\boldsymbol{\eta} \boldsymbol{\eta}^t}{\|\boldsymbol{\eta}\|_2^2}\right)\boldsymbol{X}_g|\dots|\boldsymbol{0}_n\right] \right) \\
    &= c\left(\left[\boldsymbol{x}_1|\dots|\boldsymbol{0}_n|\dots|\boldsymbol{x}_p\right] + \left[\boldsymbol{0}_n|\dots|\boldsymbol{\Pi}^\perp_{\boldsymbol{\eta}} \boldsymbol{X}_g + \frac{\boldsymbol{\eta \eta}^t}{\|\boldsymbol{\eta}\|_2}\boldsymbol{X}_g|\dots|\boldsymbol{0}_n\right] \right) \\
    &= c\left(\left[\boldsymbol{x}_1|\dots|\boldsymbol{0}_n|\dots|\boldsymbol{x}_p\right] + \left[\boldsymbol{0}_n|\dots|\boldsymbol{\Pi}^\perp_{\boldsymbol{\eta}} \boldsymbol{X}_g + \frac{\boldsymbol{\eta} \phi_g}{\|\boldsymbol{\eta}\|_2^2}|\dots|\boldsymbol{0}_n\right] \right)\qquad\mbox{with } \phi_g = {\boldsymbol{X}_g}^t\boldsymbol{\eta}  \\
    &= c\left(\left[\boldsymbol{x}_1|\dots|\boldsymbol{0}_n|\dots|\boldsymbol{x}_p\right] + \left[\boldsymbol{0}_n|\dots|\boldsymbol{\Pi}^\perp_{\boldsymbol{\eta}} \boldsymbol{X}_g + \frac{\boldsymbol{\eta} \phi_g}{\|\boldsymbol{\eta}\|_2^2}|\dots|\boldsymbol{0}_n\right] \right) \\
    &= c\left(\left[\boldsymbol{x}_1|\dots|\boldsymbol{x}_g - \frac{\boldsymbol{\eta \eta}^t \boldsymbol{x}_g}{\|\boldsymbol{\eta}\|_2^2}+ \frac{\boldsymbol{\eta} \phi_g}{\|\boldsymbol{\eta}\|_2^2}|\dots|\boldsymbol{x}_p\right]\right)
\end{align*}

We also have : $$\boldsymbol{X}_g^t\boldsymbol{\eta} \perp \boldsymbol{\Pi}^\perp_{\boldsymbol{\eta}}\boldsymbol{X}_g$$
because $\boldsymbol{\Pi}^\perp_{\boldsymbol{\eta}}$ is the orthogonal projection matrix onto the subspace orthogonal to $\operatorname{span}(\boldsymbol{\eta})$ \citep{jewell2019testing, gao2020selective}.

Finally, we have : 
\begin{align*}
    \boldsymbol{X}_g \sim \mathcal{N}_n(\boldsymbol{\mu_g}, \sigma^2_g\boldsymbol{I}_n)& \Rightarrow \boldsymbol{X}_g^t\boldsymbol{\eta} \sim \mathcal{N}(\boldsymbol{\mu}_g^t\boldsymbol{\eta}, \sigma^2_g\|\boldsymbol{\eta}\|_2^2) \\
    &\Rightarrow \phi_g \stackrel{H_0}{\sim} \mathcal{N}\left(0, \sigma^2_g\|\boldsymbol{\eta}\|^2_2\right)
\end{align*}

Thus, the p-value (\ref{eq:pvalue2}) is equal to : 
\begin{equation*}
    \mathbb{P}_{H_0}\left(|\phi_g|>|{\boldsymbol{X}_g}^t\boldsymbol{\eta}| \mbox{ | }\phi_g \in S_g \right)
\end{equation*}
with $S_g = \left\{\phi_g : C_k, C_l \in c\left(\boldsymbol{X}(\phi_g)\right)\right\}$ and $\phi_g \stackrel{H_0}{\sim} \mathcal{N}\left(0, \sigma^2_g\|\boldsymbol{\eta}\|^2_2\right)$
\newpage

\section{Numerical computation of the selective p-value}

Since the selective p-value given in \eqref{eq:pvalue2} is intractable, in practice it is computed using a Monte-Carlo approach with important sampling resulting in the p-value described in \eqref{eq:pvalueImpSamp}. However, \citet{phipson2010permutation} showed that the classical Monte-Carlo estimator of a p-value could be biased for near to zero p-values. In fact, very tiny p-values could be approximated by exactly $0$ using the Monte-Carlo approach, leading to statistical hypothesis testing that does not efficiently control the type I error rate. To overcome this problem, they propose to correct the Monte-Carlo p-value by adding $1$ in both the numerator and the denominator of the estimated p-value. With this correction, instead of having exactly $0$ Monte-Carlo p-value, the near to zero p-values are approximated by $\frac{1}{N+1}$ where $N$ is the number of Monte-Carlo samples. 

Unfortunately, we showed using numerical simulation studies that this correction could not work with the selective p-value computed as in \eqref{eq:pvalueImpSamp} for two reasons. The first one is because this p-value originally come from a conditional probability \eqref{eq:pvalue2}, so by definition, there are in fact two probabilities to compute and to correct. The second problem arrives because of the important sample approach. In fact, because under $\mathcal{H}_1$ the $\pi^i$ in \eqref{eq:pvalueImpSamp} are very small, adding $1$ will drastically change the scale of the p-value. 
\noindent So, because of the important sampling approach, we need to correct our Monte-Carlo p-value by adding a constant in the same order of $\pi$. We propose to add $\overline{\pi} = \frac{1}{N}\sum\limits_{i=1}^N \pi^i\mathbb{1}\left\{C_k, C_l \in c(\matrice{X}(\omega_g ^i))\right\}$ in both the numerator and the denominator of \eqref{eq:pvalueImpSamp}. 
This correction is reasonable since for small p-value, that is under $\mathcal{H}_1$ we have :

\begin{enumerate}
    \item [i)]$|\matrice{x}_g^t\matrice{\eta}|$ is large because $C_k$ and $C_l$ are truly separated on $\matrice{X}_g$
    \item [ii)] $\sum\limits_{i=1}^N \pi^i\mathbb{1}\{|\omega_g^i| > |\boldsymbol{x}_g^t\boldsymbol{\eta}|, C_k, C_l \in c(\boldsymbol{X}(\omega_g^i)\} \simeq 0$ since each $\pi^i = \frac{f_1(\omega^i_g)}{f_2(\omega^i_g)}$ where $f_1$ is the density of a gaussian distribution with mean $0$. Then, because $\omega^i_g \sim \mathcal{N}(\boldsymbol{x}_g^t\boldsymbol{\eta}, \sigma^2\|\boldsymbol{\eta}\|_2)$ where $|\matrice{x}_g^t\matrice{\eta}|$ is large, $f_1$ is evaluated in a point that is far away of the mean, and that is why $f_1(\omega^i_g)\simeq 0$.
\end{enumerate}

\noindent So using i) and ii): 

\begin{align*}
    \frac{\sum\limits_{i=1}^N \pi^i \mathbb{1}\left\{|\omega_g^i| \geq |\matrice{x}_g^t\matrice{\eta}|, C_k, C_l \in c(\matrice{X}(\omega_g^i))\right\} + \overline{\pi}}{\sum\limits_{i=1}^N \pi^i \mathbb{1}\left\{C_k, C_l \in c(\matrice{X}(\omega_g ^i))\right\} + \overline{\pi}} &=
     \frac{0 + \overline{\pi}}{\sum\limits_{i=1}^N \pi^i \mathbb{1}\left\{C_k, C_l \in c(\matrice{X}(\omega_g ^i))\right\} + \overline{\pi}} \\
     &\simeq \frac{\frac{1}{N}\sum\limits_{i=1}^N \pi^i\mathbb{1}\left\{C_k, C_l \in c(\matrice{X}(\omega_g ^i))\right\}}{\sum\limits_{i=1}^N \pi^i \mathbb{1}\left\{C_k, C_l \in c(\matrice{X}(\omega_g ^i))\right\} +\frac{1}{N}\sum\limits_{i=1}^N \pi^i\mathbb{1}\left\{C_k, C_l \in c(\matrice{X}(\omega_g ^i))\right\}}  \\
     &= \frac{\frac{1}{N}\sum\limits_{i=1}^N \pi^i\mathbb{1}\left\{C_k, C_l \in c(\matrice{X}(\omega_g ^i))\right\}}{\sum\limits_{i=1}^N \pi^i \mathbb{1}\left\{C_k, C_l \in c(\matrice{X}(\omega_g ^i))\right\}\left[1+\frac{1}{N}\right]} \\
     &= \frac{\frac{1}{N}}{1+\frac{1}{N}} \\
     &= \frac{1}{N+1}
\end{align*}

\noindent So, by correcting our Monte-Carlo p-value by adding $\overline{\pi}$ we obtain the estimator proposed by \citet{phipson2010permutation} for small p-value. 

\newpage
\section{Supplementary Figure 1}
\begin{figure}[h!]
    \centering
    \includegraphics[width = 0.75\textwidth]{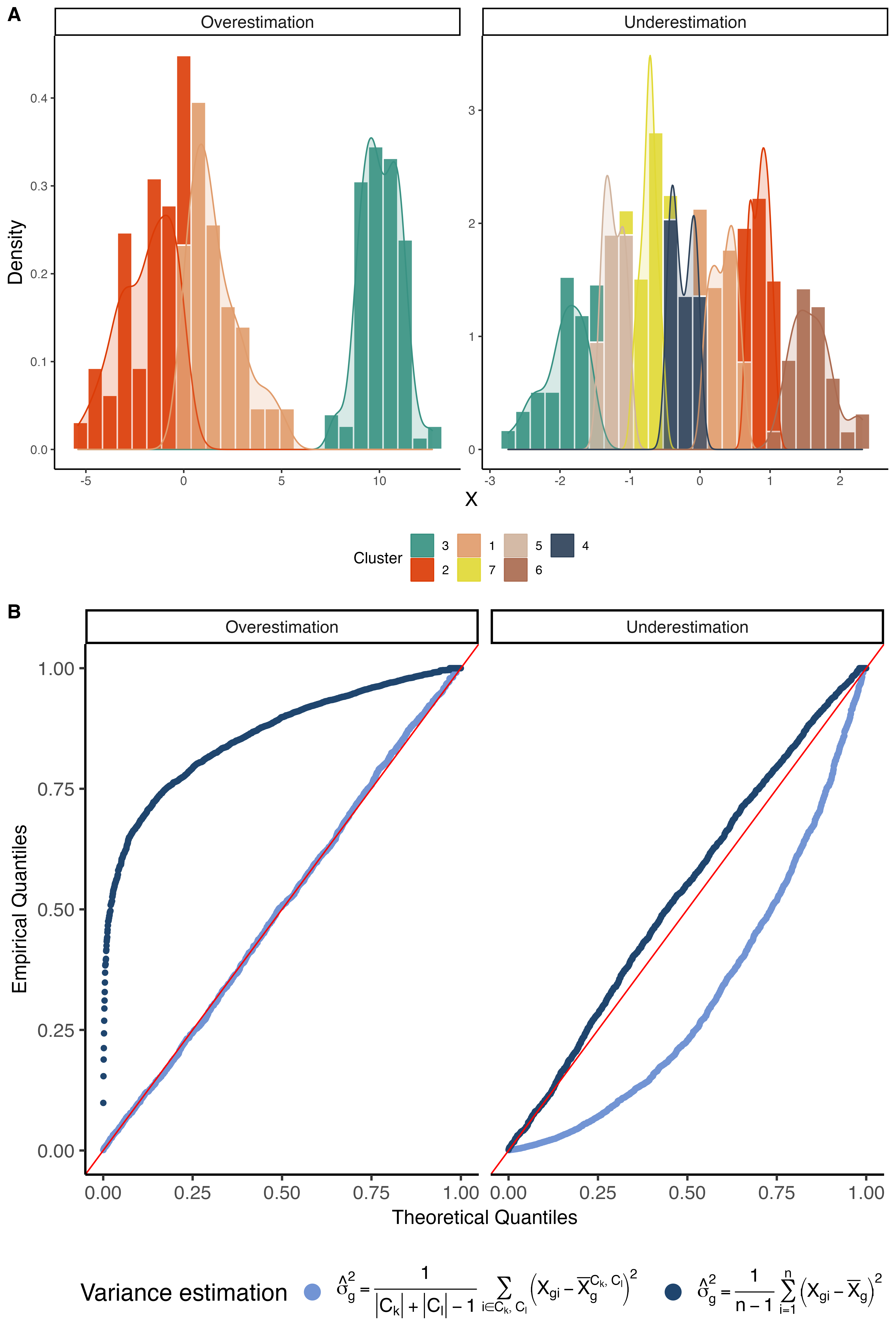}
    \caption{Impact of the variance estimation on the p-values of the selective test. \textbf{panel A} The data are simulated according to a two-component Gaussian mixture: $X \sim 0.5\mathcal{N}\left(0, 4\right) + 0.5\mathcal{N}(10,1)$ for the overestimation panel and according to a standard Gaussian distribution with mean $0$ and variance $1$ for the underestimation panel. \textbf{panel B} QQ-plot of selective p-values against the uniform distribution according to the variance estimator for the test Cluster 1 vs. Cluster 2 for $2000$ simulations of the data. Overestimation of variance leads to conservative p-values, while underestimation leads to false positives.}
    \label{fig:S1}
\end{figure}
\newpage

\section{Supplementary Figure 2}
\begin{figure}[htbp]
    \centering
    \includegraphics[width =\textwidth]{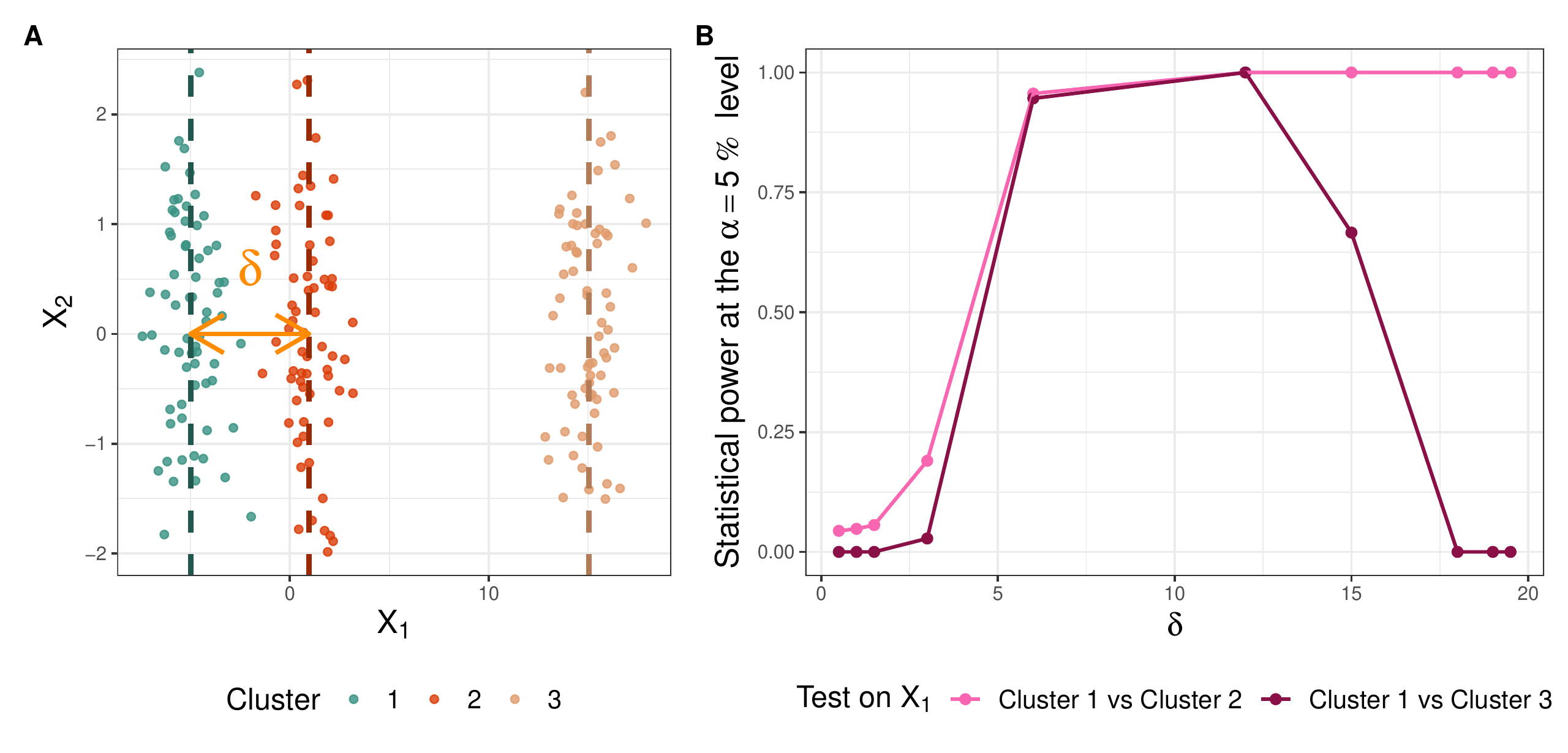}
    \caption{Illustration of the possible loss of statistical power of the selective test in cases where there are more than two estimated clusters. \textbf{panel A} Data generation process. A bivariate dataset is simulated such as three clusters are all separated only on $X_1$. Cluster 1 and Cluster 2 are separated according to a mean difference $\delta \in \{0.5, 1, 1.5, 3, 6, 12, 15, 18, 19, 19.5\}$. \textbf{panel B} Statistical power at the $\alpha = 5\%$ level of the selective test computed using $500$ simulation of the data as described in \textbf{panel A} according to $\delta$, the mean difference between Cluster 1 and Cluster 2. For each simulation, the selective test is applied to test the separation of Cluster 1 vs Cluster 2 and Cluster 1 vs Cluster 3 only on $X_1$.}
    \label{fig:S2}
\end{figure}
\newpage
\section{Supplementary Figure 3}

\begin{figure}[htbp]
    \centering
    \includegraphics[width =\textwidth]{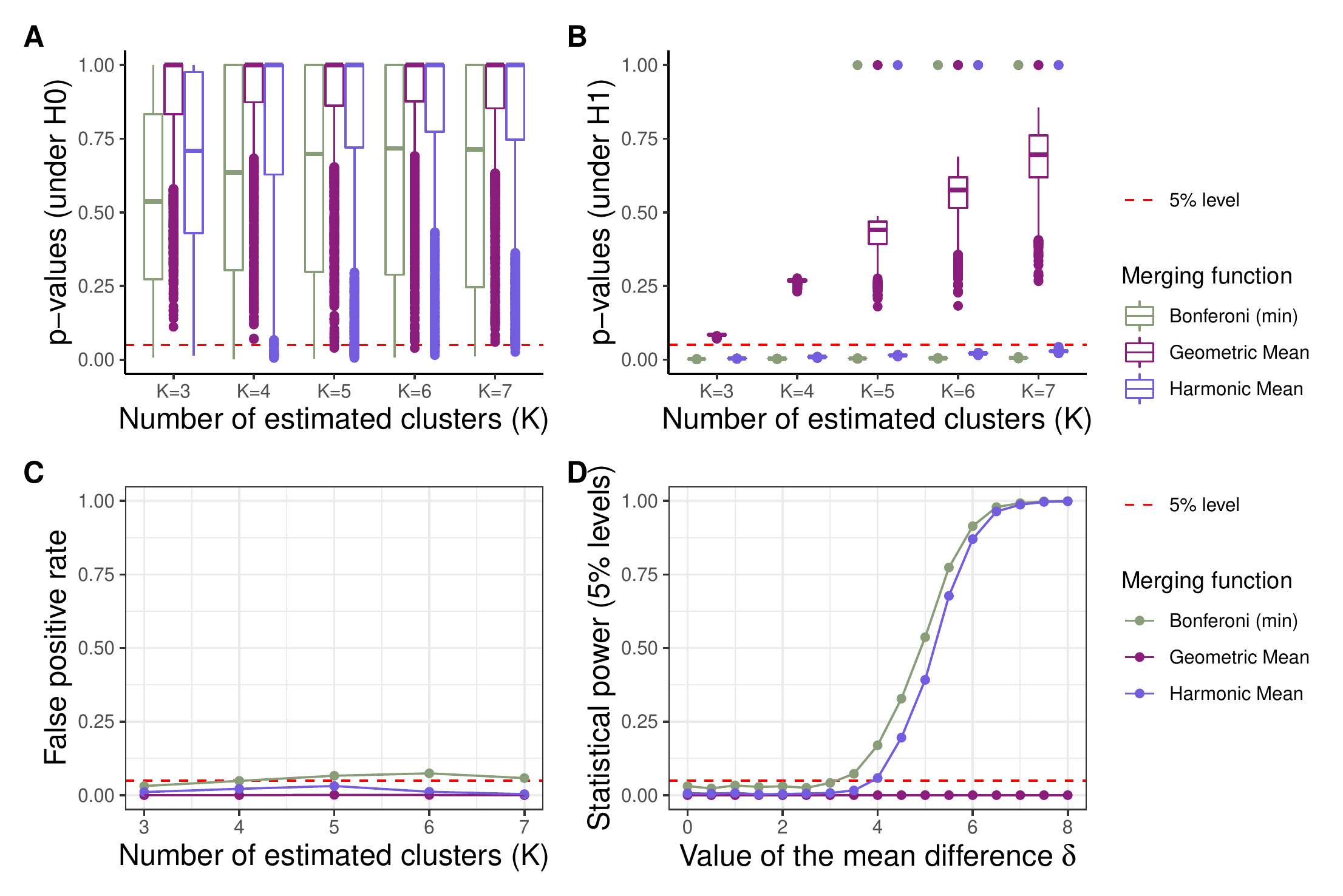}
    \caption{Comparison of three different merging functions presented in \citet{vovk2020combining}. \textbf{panel A} Distribution of the resulting merging p-values under $H_0$ as a function of the number of estimated clusters. \textbf{panel B} Distribution of the resulting merging p-values under $H_1$ (Gaussian mixture with only two components of equal proportion and variance) as a function of the number of estimated clusters. \textbf{panel C} False positive rate as a function of the number of estimated clusters. \textbf{panel D} Statistical power as a function of the mean difference $\delta$ between the two modes of the mixture where $K=4$ clusters are estimated (the same simulation as in Figure 3. The selective test is always applied to the most extreme clusters, and in such a way that the maximum number of adjacent p-values are merged. $2000$ simulations of the data were used.}
    \label{fig:S3}
\end{figure}
\newpage
\section{Supplementary Figure 4}

\begin{figure}[htbp]
    \centering
    \includegraphics[width = \textwidth]{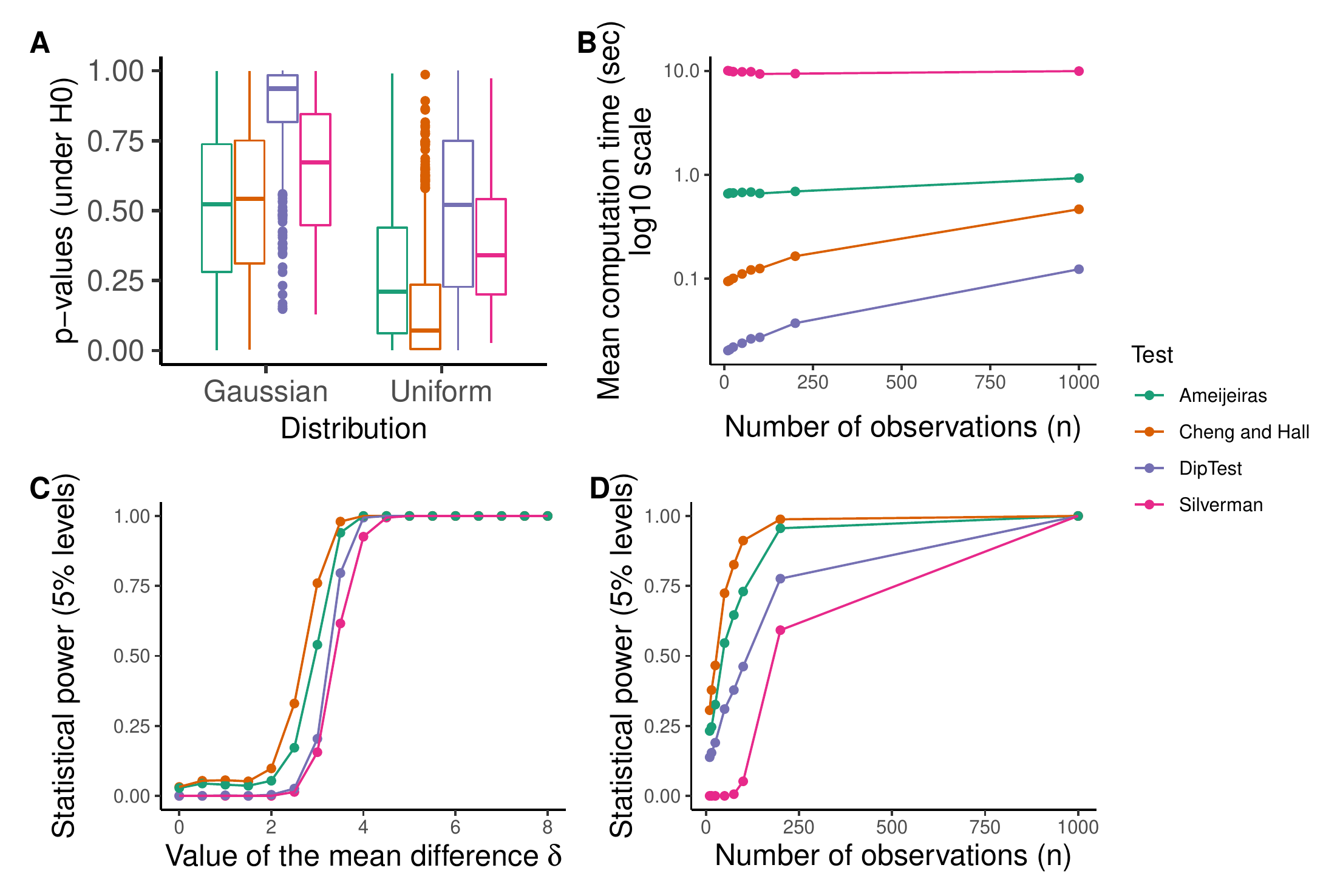}
    \caption{Comparison of different multimodality tests implemented in the \texttt{R} package \texttt{multimode}\citep{ameijeiras2021multimode}. \textbf{panel A} p-values of each multimodality tests under the null for $500$ simulations of $200$ realisations of the Gaussian and uniform distributions. \textbf{panel B} Mean computation time required by each test as a function of the number of observations $n$ (averaging over the $500$ simulations. \textbf{panel C} Statistical power (at the $\alpha = 5\%$ level) of each multimodality test as a function of $\delta$, the difference in means between two modes of a two-components Gaussian mixture ($n=200$ observations). \textbf{panel D} Statistical power( at the $\alpha = 5\%$ level) of each multimodality test as a function of the number of observations for $\delta = 3.5$ fixed.}
    \label{fig:S4}
\end{figure}
\newpage
\section{Supplementary Figure 5}
\begin{figure}[htbp]
    \centering
    \includegraphics[width = \textwidth]{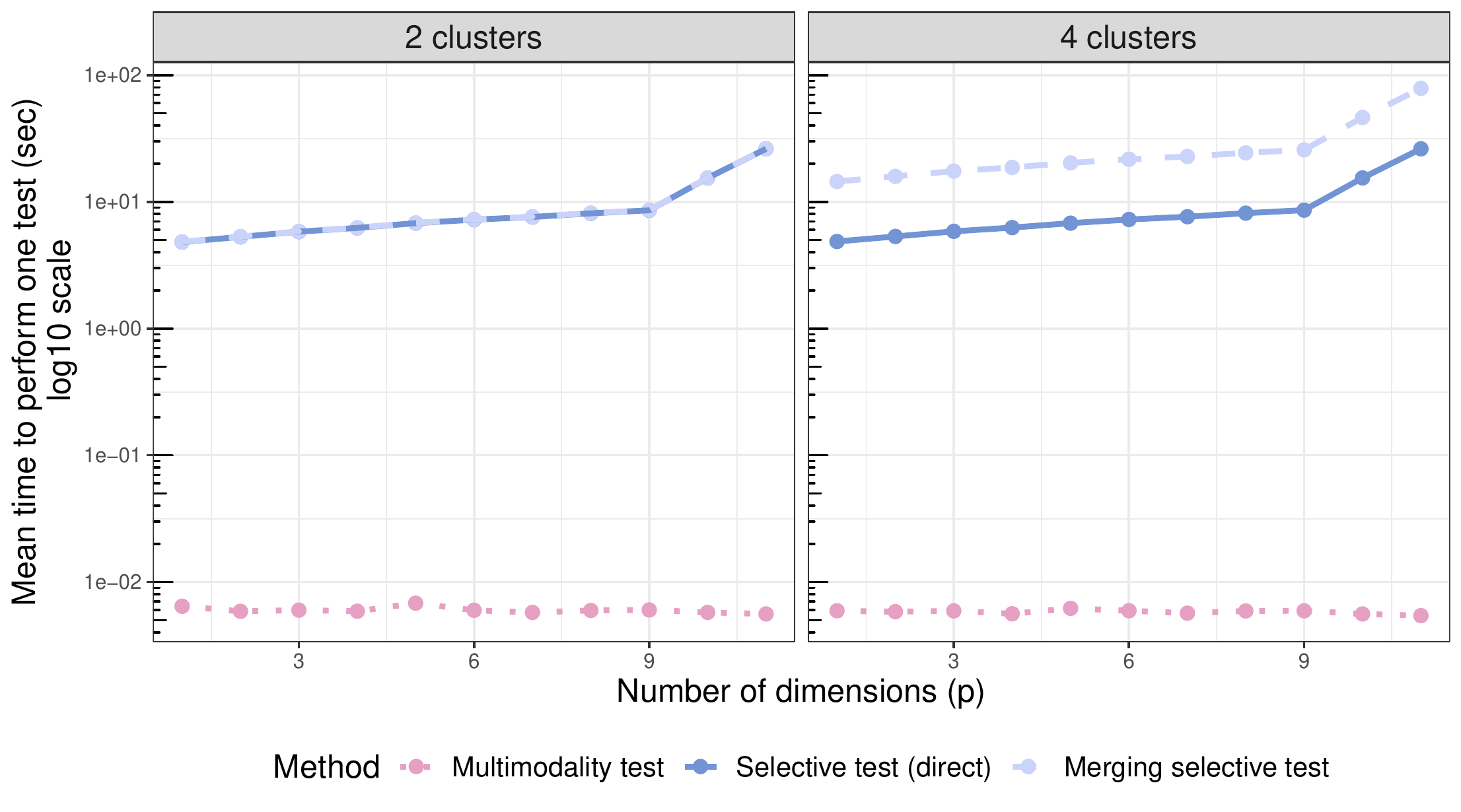}
    \caption{Mean computational time of the three proposed tests (based on $500$ simulations of the data) according to the number of dimensions (p) of $\matrice{X}$. The tests are performed only for the first variable, so the dimensionality of the data only affects the computation times of the selective tests since the clustering method must be applied on the data for each Monte-Carlo simulation using all the dimensions of $\matrice{X}$}
    \label{fig:S5}
\end{figure}
\newpage

\section{Supplementary Figure 6}
\begin{figure}[htbp]
    \centering
    \includegraphics[width = \textwidth]{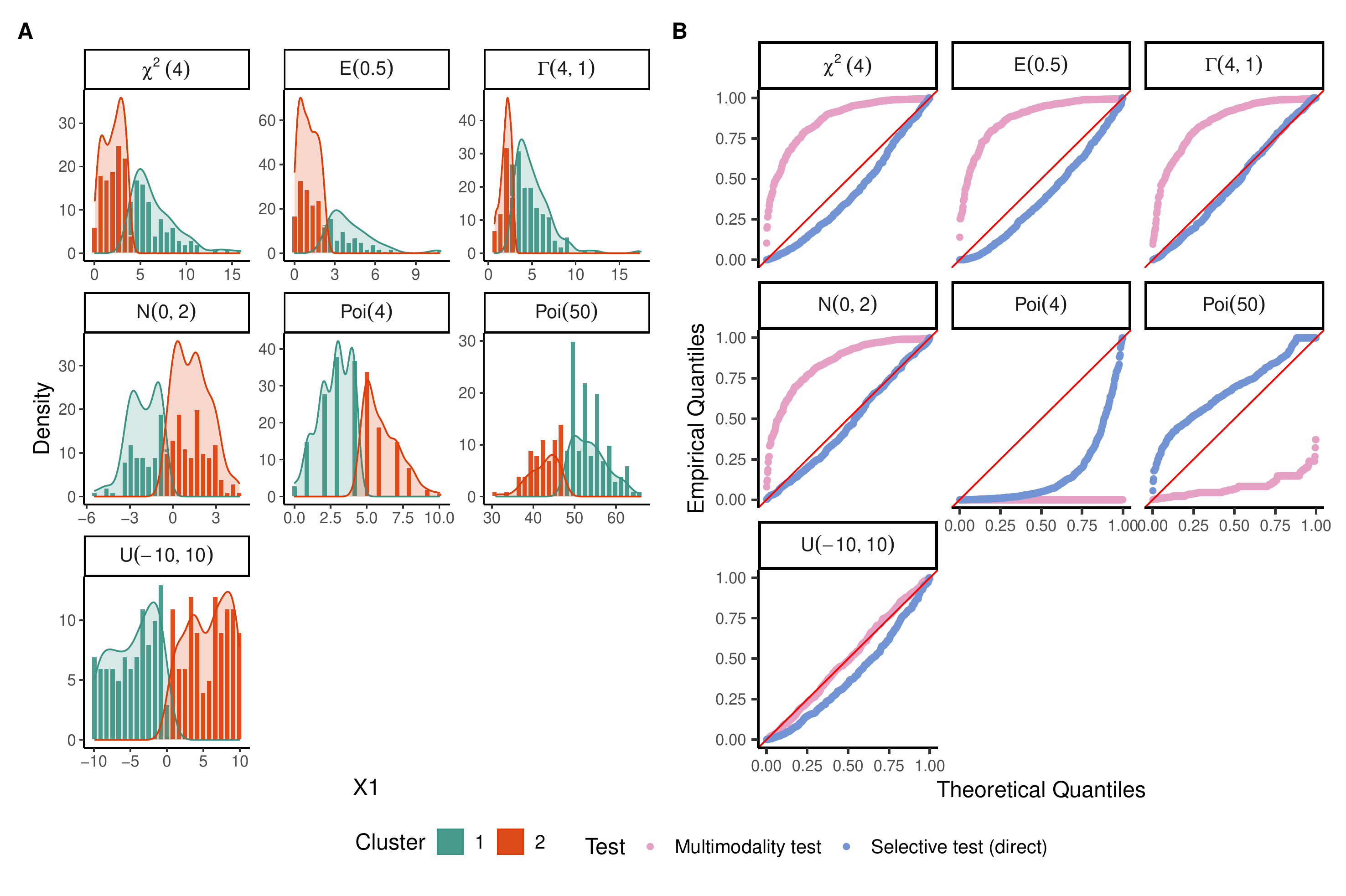}
    \caption{Robustness of our tests for misspecification of the distributional assumption under the null. \textbf{panel A} Data generation. The behaviour of our tests was studied on $7$ different univariate and unimodal distributions. For each distribution, Ward's clustering on euclidean distance was applied to build $2$ clusters. \textbf{panel B} QQ-plot against the Uniform distribution of the p-values returned by our two tests for $500$ simulations of each distribution.}
    \label{fig:S6}
\end{figure}
  
\newpage

\section{Supplementary Table 1}
\begin{table}[ht]
\centering
\begin{tabular}{lcccc}
  \hline
\textbf{Cluster pair} & \textbf{bill length} & \textbf{bill depth} & \textbf{flipper length} & \textbf{body mass} \\
  \hline
Cluster 1 vs Cluster 2 & 1.67 & 1.53 & 1.94 & 1.75 \\
  Cluster 1 vs Cluster 3 & 0.16 & 1.93 & 0.50 & 0.16 \\
  Cluster 2 vs Cluster 3 & 1.83 & 0.40 & 1.44 & 1.59 \\
   \hline
\end{tabular}
\caption{Values of the mean difference ($\delta$) on each (scaled) variable between each estimated pair of clusters}
\label{tab:delta_appli}
\end{table}
\newpage
\end{document}